\documentstyle[prd,aps,floats,epsf]{revtex}

\begin{document}
\twocolumn[\hsize\textwidth\columnwidth\hsize\csname
@twocolumnfalse\endcsname

\draft
\title{\vspace*{-1cm}\hfill\mbox{\small FSU-SCRI-99-09}\\
\vspace*{0.6cm}Small eigenvalues of the staggered Dirac operator in the adjoint
representation and Random Matrix Theory}
\author{
Robert G. Edwards and Urs M. Heller}
\address{
SCRI, Florida State University, 
Tallahassee, FL 32306-4130, USA}
\author{ Rajamani Narayanan}
\address{Dept. of Physics, Bldg. 510A,
Brookhaven National Laboratory,
P. O. Box 5000, Upton, NY 11973}

\maketitle 

\begin{abstract}
The low-lying spectrum of the Dirac operator is predicted to be universal,
within three classes, depending on symmetry properties specified according
to random matrix theory. The three universal classes are 
the orthogonal, unitary and symplectic ensemble.
Lattice gauge theory with staggered fermions has verified two of the cases
so far, unitary and symplectic, with staggered fermions in the fundamental
representation of SU(3) and SU(2). We verify the missing case here, namely
orthogonal, with staggered fermions in the adjoint representation of
SU($N_c$), $N_c=2, 3$.
\end{abstract}
\pacs{PACS numbers: 11.15.Ha, 11.30Rd, 12.38Gc}
]

Random matrix theory (RMT) has been successful in predicting spectral
properties of QCD-like theories in the so-called microscopic scaling regime,
defined by $1/\Lambda_{QCD} << L << 1/m_\pi$, with $L$ the linear
extent of the system~\cite{SV,JV94}. Assuming spontaneous chiral symmetry
breaking in the underlying theory, this is the regime dominated by the soft
pions associated with the chiral symmetry breaking~\cite{LS}. Up to a scale,
given by the chiral condensate (at infinite volume) $\Sigma = \langle
\bar \psi \psi \rangle$, the distribution of the low lying eigenvalues,
the so-called microscopic spectral density
\begin{equation}
\rho_S(z) = \lim_{V \to \infty} \frac{1}{V} \rho \left(
 \frac{z}{V\Sigma} \right)
\label{eq:rho_S}
\end{equation}
with $\rho(\lambda)$ the usual (macroscopic) spectral density
and, in particular, the rescaled distribution of the lowest
eigenvalue $P_{\rm min}(z)$ with $z = V\Sigma \lambda_{\rm min}$ are
universal, dependent only on the symmetry properties, the number of
dynamical quark flavors and the number of exact zero modes, {\it i.e.,}
the topological sector, but not the potential in RMT~\cite{univ}.
Recently, these properties have been derived directly from the effective,
finite-volume partition functions of QCD of Leutwyler and Smilga, without
the detour through RMT~\cite{noRMT}.

The lattice fermion action should have a chiral symmetry for the
predictions of random matrix theory to apply.
Until recently, staggered fermions were the only fermions regularized
on the lattice that retained a chiral symmetry\footnote{A new lattice
regularization of massless fermions with good chiral properties, and
even an index theorem, has recently been developed~\cite{herbert}.
The RMT predictions for these overlap fermions have been verified for
examples in all three ensembles, including the classification into
different topological sectors, in~\cite{ehkn}.}.
The RMT predictions have
been nicely verified for staggered fermions in the fundamental
representations of SU(2)~\cite{SU2_q} and SU(3)~\cite{SU3_q} gauge group
in quenched QCD, and for SU(2) also with dynamical fermions~\cite{SU2_d}.
These represent two out of the three different cases
predicted by RMT, chiral symplectic and chiral unitary. Staggered fermions
are not always in the same universality class as continuum fermions
in the context of random matrix theory. Staggered and continuum fermions
belong to the unitary ensemble when the fermions are in the fundamental
representation of SU($N_c$) doe $N_c \ge 3$.
But staggered fermions in the fundamental 
representation of SU(2) belong to the symplectic ensemble since their
entire spectrum is two-fold degenerate~\cite{HV95} in contrast
to continuum fermions which belong to the orthogonal ensemble~\cite{JV94}.
Another example where the staggered
fermions and continuum fermions are not in the same universality class
is when the fermions are in the adjoint representation of SU($N_c$).
Here the situation is just reversed: continuum fermions are in the
symplectic ensemble~\cite{JV94} with a two-fold degeneracy of the entire
spectrum\footnote{To compare with RMT predictions only one of each degenerate
pair of eigenvalues is kept and the adjoint fermion is thereby considered
as a Majorana fermion.}
while staggered fermions are in the orthogonal ensemble -- they are
real, since adjoint gauge fields are real and since the ``Dirac matrices''
for staggered fermions are just phases, $\pm 1$. 
We therefore use this example with staggered fermions in the adjoint
representation of SU($N_c$) to test the missing case.

In this report, we consider staggered fermions in the adjoint representation
of SU(2) and SU(3) in the quenched approximation. Since staggered fermions
(at finite lattice spacing) do not obey an index theorem and thus have
no exact zero modes~\cite{KSzeromode}, the RMT predictions for $\nu=0$
are expected to apply. 
This has been found in the studies with staggered fermions
in the fundamental representation~\cite{SU2_q,SU3_q,SU2_d}, and our results
will support it for staggered fermions in the adjoint representation.
The rescaled distribution of the lowest eigenvalue, $z= V\Sigma
\lambda_{\rm min}$, should thus be \cite{Forrester}
\begin{equation}
P_{\rm min}(z) = {2+z\over 4} e^{-{z\over2}-{z^2\over 8}} ~.
\label{eq:pmin_OE}
\end{equation}
This distribution is quite distinct from all other cases: it starts at a
finite, non-zero value at $z=0$. Given the chiral condensate, $\Sigma$,
(\ref{eq:pmin_OE}) is a parameter free prediction. If $\Sigma$ is not
otherwise known, it can be determined from a one-parameter fit of the
distribution of the lowest eigenvalue to (\ref{eq:pmin_OE}).

\begin{figure*}[t]
\vspace*{-10mm} \hspace*{-0cm}
\begin{center}
\epsfxsize = 0.8\textwidth
\centerline{\epsfbox[100 175 550 500]{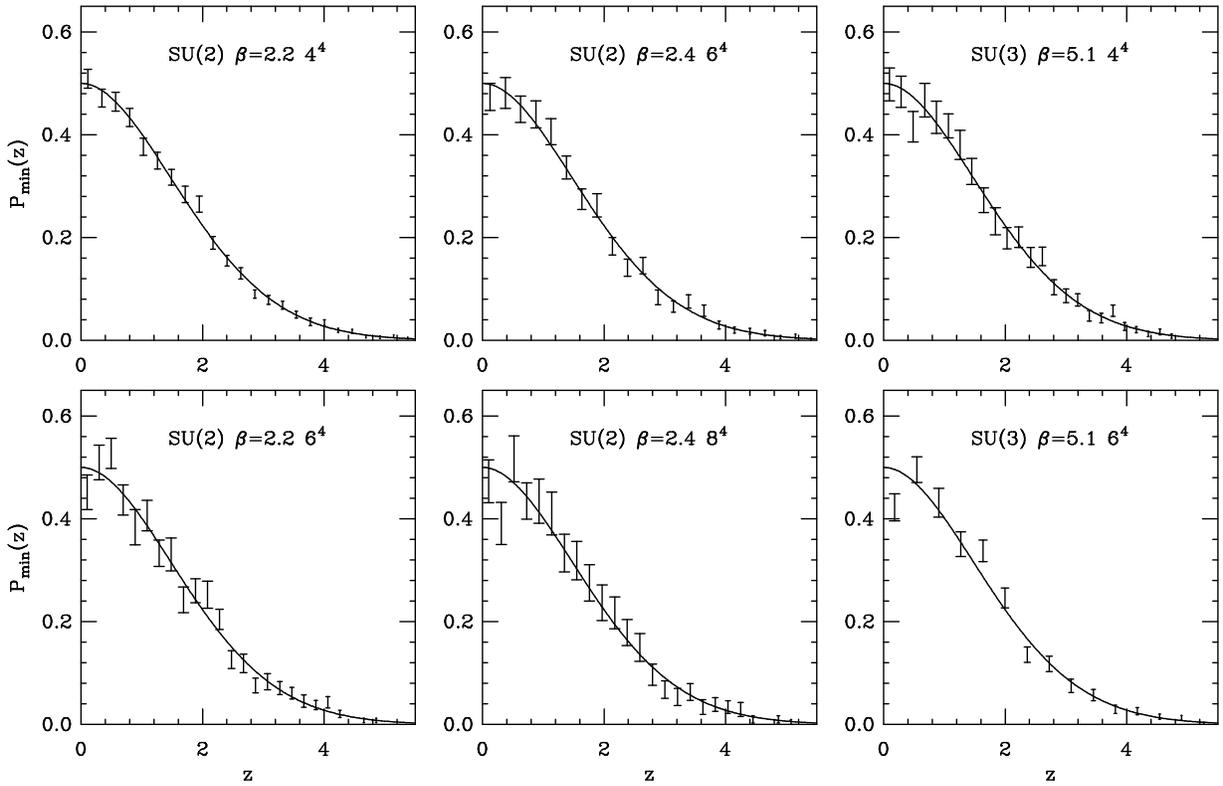}}
\end{center}
\caption{
The distribution of the rescaled lowest eigenvalue, $P_{\rm min}(z)$,
together with the RMT prediction using the best estimate for the
chiral condensate.}
\label{fig:pmin}
\end{figure*}

We have computed the low lying eigenvalues of the staggered Dirac operator
in the adjoint representation with the Ritz functional method~\cite{Ritz},
applied to $D^\dagger D = - D^2$, for several SU(2) gauge field ensembles
and for two SU(3) ensembles. The distribution of the lowest eigenvalue,
approximated by a histogram with jackknife errors, was fit to the
predicted form, (\ref{eq:pmin_OE}), with $\Sigma$, the infinite volume
value of $\langle \bar \psi \psi \rangle$ as the only free parameter.
The results of the fit, together with the number of configurations
considered, gauge coupling and lattice size, are given in Table~\ref{tab:Sigma}.
In all cases we obtained good fits to the predicted form. For most gauge
couplings we considered two different lattice sizes. The values for
$\Sigma$ obtained from the two different lattice sizes agree well, within
statistical errors. Some distributions, together with the fitted
analytical predictions, are shown in Figure~\ref{fig:pmin}.

\begin{table}
\caption{The chiral condensate, $\Sigma$, from fits of the distribution
of the lowest eigenvalue to the RMT predictions. The last column gives
the confidence level of the fit.}
\label{tab:Sigma}
\begin{tabular}{|c|c|c|c|c|c|}
\hline
 Group & $\beta$ & $L$  & $N$ & $\Sigma$ & $Q$ \\
\hline
 SU(2) & 1.8 & 4 & 5000 & 1.796(18) & 0.921 \\
 SU(2) & 2.0 & 4 & 5000 & 1.676(17) & 0.640 \\
 SU(2) & 2.0 & 6 & 2000 & 1.722(30) & 0.496 \\
 SU(2) & 2.2 & 4 & 5000 & 1.489(16) & 0.840 \\
 SU(2) & 2.2 & 6 & 2000 & 1.529(28) & 0.344 \\
 SU(2) & 2.4 & 6 & 2000 & 1.212(21) & 0.852 \\
 SU(2) & 2.4 & 8 & 1000 & 1.264(30) & 0.874 \\
\hline
 SU(3) & 5.1 & 4 & 2500 & 4.724(76) & 0.786 \\
 SU(3) & 5.1 & 6 & 1500 & 4.671(91) & 0.035 \\
\hline
\end{tabular}
\end{table}

\begin{figure*}[t]
\vspace*{-10mm} \hspace*{-0cm}
\begin{center}
\epsfxsize = 0.8\textwidth
\centerline{\epsfbox[100 175 550 500]{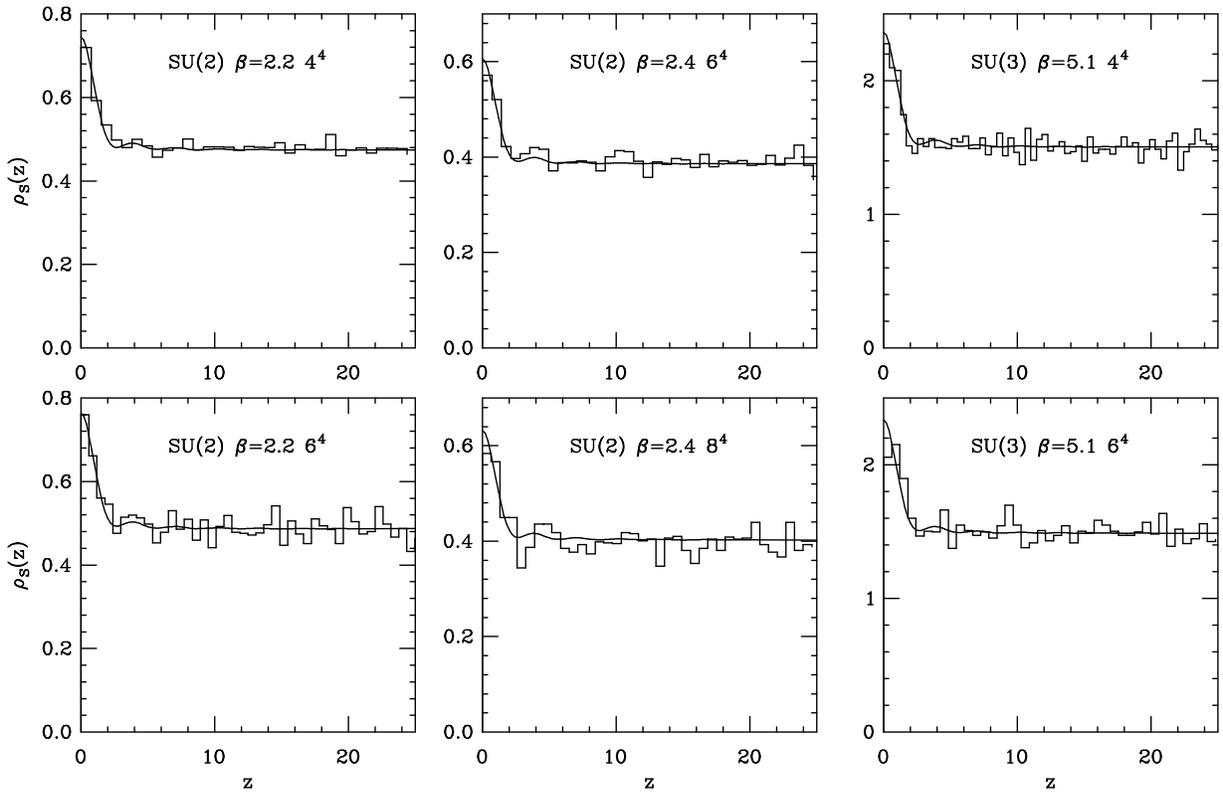}}
\end{center}
\caption{
The microscopic spectral density, $\rho_S$, together with the RMT
prediction~\protect\cite{JVsu2} using the best estimate for the
chiral condensate.}
\label{fig:rho_S}
\end{figure*}

The consistency of the extracted value for the chiral condensate, $\Sigma$,
from the two different lattice sizes makes it evident that we could have
used the value obtained from one lattice size to get a parameter free
description of the results from the other lattice size.
Given the scale $\Sigma$ the microscopic spectral density
$\rho_S$, eq.~(\ref{eq:rho_S}), is predicted by results of RMT
\cite{JVsu2}. The comparison for the same systems as in Figure~\ref{fig:pmin}
is shown in Figure~\ref{fig:rho_S}. The agreement is quite nice, extending
over the entire region covered by the eigenvalues that we determined.
Note that for the chiral orthogonal ensemble the oscillations in the
RMT prediction for $\rho_S$, except the first one coming from the lowest
eigenvalue, are very small. Obviously, we would need much more
statistics to resolve additional ``wiggles'' in our data.

In conclusion, adjoint staggered fermions are argued to belong to the
chiral orthogonal ensemble of RMT. We found that the RMT predictions
indeed describe the spectrum of low lying eigenvalues of the staggered
Dirac operator in the adjoint representation very well for both SU(2)
and SU(3) in our quenched simulations.

\acknowledgements
This research was supported by DOE contracts 
DE-FG05-85ER250000 and DE-FG05-96ER40979.
Computations were performed on the workstation cluster at SCRI.
We thank P.~Damgaard for discussions and J. Verbaarschot for making
the data for the analytical curve in Figure~\ref{fig:rho_S} available.

\end{document}